# Superconductivity in infinite-layer nickelate $La_{1-x}Ca_xNiO_2$ thin films

S. W. Zeng[1,#,*], C. J. Li[2,3,#], L. E. Chow[1,#], Y. Cao[4], Z. T. Zhang[1], C. S. Tang[5,6], X. M. Yin[7], Z. S. Lim[1], J. X. Hu[1], P. Yang[2,5], A. Ariando[1,*]

[1]*Department of Physics, Faculty of Science, National University of Singapore, Singapore 117551, Singapore*

[2]*Department of Materials Science and Engineering, National University of Singapore, Singapore 117575, Singapore*

[3]*Department of Materials Science and Engineering, Southern University of Science and Technology, Shenzhen 518055, China*

[4]*Department of Electrical and Computer Engineering, National University of Singapore, 4 Engineering Drive 3, Singapore 117583, Singapore*

[5]*Singapore Synchrotron Light Source (SSLS), National University of Singapore, Singapore 117603, Singapore*

[6]*Institute of Materials Research and Engineering, A\*STAR (Agency for Science, Technology and Research), 2 Fusionopolis Way, Singapore 138634, Singapore*

[7]*Shanghai Key Laboratory of High Temperature Superconductors, Physics Department, Shanghai University, Shanghai 200444, China*

[#]The authors contributed equally to this work.

[*]To whom correspondence should be addressed.

E-mail:

ariando@nus.edu.sg; shengwei_zeng@u.nus.edu

**Abstract**

We report the observation of superconductivity in infinite-layer Ca-doped LaNiO$_2$ (La$_{1-x}$Ca$_x$NiO$_2$) thin films and construct their phase diagram. Unlike the metal-insulator transition in Nd- and Pr-based nickelates, the undoped and underdoped La$_{1-x}$Ca$_x$NiO$_2$ thin films are entirely insulating from 300 K down to 2 K. A superconducting dome is observed at $0.15 < x < 0.3$ with weakly insulating behavior at the overdoped regime. Moreover, the sign of the Hall coefficient $R_H$ changes at low temperature for samples with a higher doping level. However, distinct from the Nd- and Pr-based nickelates, the $R_H$-sign-change temperature remains at around 35 K as the doping increases, which begs further theoretical and experimental investigation in order to reveal the role of the $4f$ orbital to the (multi)band nature of the superconducting nickelates. Our results also emphasize the significant role of lattice correlation on the multiband structures of the infinite-layer nickelates.

**Introductions**

The resemblance of the high-$T_c$ superconducting cuprates in the nickel-based compounds has made superconductivity in the nickelates to be one of the holy grails in superconducting research in the last three decades. The recent discovery of superconductivity in infinite layer Sr-doped NdNiO$_2$ (*1*) has now further propelled the importance of nickelates as a route towards understanding and a material platform for searching for new high-$T_c$ superconductors (*1-6*). The perovskite nickelates $R$NiO$_3$ ($R$ = rare-earth element) show a structural transition and exhibit a metal-insulator transition upon cooling when $R$ = Nd and Pr, while it remains metallic when $R$ = La (7). To mimic the cuprate-like $3d^9$ electronic configuration, the infinite-layer $LaNiO_2$ is required and can be realized by oxygen deintercalation of the $LaNiO_3$ precursor (8). Much effort has been devoted to dope carriers in this $LaNiO_2$, but it failed to obtain superconductivity (1). To increase the electronic bandwidth, the La was substituted by a rare earth element with a smaller ionic radius (e.g., Nd atom), resulting in the observation of superconductivity in $Nd_{0.8}Sr_{0.2}NiO_2$ (1). This route to superconductivity contrasts the heavily studied copper- and iron-based superconductors in which superconductivity was first obtained in La-based compounds (9, 10), while up to now, the superconducting nickelate has only been observed in Sr-doped compounds with $R$ = Nd or Pr (i.e., $R_{1-x}Sr_xNiO_2$) (1, 11-20). Moreover, the trilayer nickelates $R_4Ni_3O_8$ ($R$ = La and Pr) that possesses the same $NiO_2$ square plane as in the infinite-layer nickelates and an effective 1/3 hole doping showed an insulating ground state in the case of $La_4Ni_3O_8$, in contrast to the metallic state in the $Pr_4Ni_3O_8$ (21). Theoretical calculation attributed the absence of superconductivity to the intercalation of H atoms in energetically favourable $LaNiO_2$ to form $LaNiO_2H$ during the reduction process (22). It has also been shown that hybridization with Nd $4f$ orbitals is a non-negligible ingredient for the superconductivity in doped $NdNiO_2$, while the $f$ orbitals are far from the Fermi level in $LaNiO_2$ (23-25). On the other hand, while earlier density functional theory (DFT) studies for $NdNiO_2$ treated the $4f$ orbitals as a "frozen core" under the assumption that the $4f$ orbitals are localized and do not show around the Fermi level (26, 27), a recent DFT combined with dynamical mean-field theory (DMFT) calculation comparing nickelates with and without $4f$ orbitals also concluded that $4f$ orbitals are not essential to superconductivity in nickelates (28). This suggests the importance of elucidating the role of $4f$ state of the electronic structure in the occurrence of superconductivity in nickelates (23-25). Therefore, the realization of superconductivity in the La series is a key step to understanding the superconducting properties in the nickelates.

In unconventional superconductors containing the rare-earth element, the substitution of an isovalent ion with a smaller radius causes an increase in inner chemical pressure due to the shrinkage of the crystal lattice. This is considered an important method to enhance $T_c$, as experimentally demonstrated in cuprates and pnictides (*29, 30*). In the Nd- and Pr-based nickelates, the $R_H$ sign-change temperature varies with Sr doping level, indicating that the Sr doping not only introduces charges but also modulates the band structures (*4, 11, 12, 15, 31-33*). Moreover, the lattice constant increases with increasing Sr doping level, and the superconducting dome occurs near the $R_H$-sign-change boundary. It is therefore anticipated to induce superconductivity in $LaNiO_2$ through doping with a smaller ionic radius. Here we present the observation of superconductivity in infinite-layer $La_{1-x}Ca_xNiO_2$ thin films through Ca doping.

**Results**

**Doping-dependent structural properties**

The perovskite $La_{1-x}Ca_xNiO_3$ precursor thin films were synthesized by a pulsed laser deposition technique and reduced to an infinite-layer phase using the soft-chemistry topotactic reduction method. The X-ray diffraction (XRD) $\theta$–$2\theta$ patterns of the as-grown $La_{1-x}Ca_xNiO_3$ thin films can be found in Supplementary Data Fig. S1. The prominent thickness oscillations in the vicinity of the (002) peak indicate the single-phase and high quality of the perovskite films. After reduction, a clear transition of the diffraction peaks is seen, confirming the transformation from the perovskite to the infinite-layer structure, as shown in Fig. 1A for a representative Ca doping level of 0.23. Figure 1B shows a high-angle annular dark-field scanning transmission electron microscopy (HAADF-STEM) image of the 17-nm $La_{0.77}Ca_{0.23}NiO_2$ thin film. A clear infinite-layer structure is observed with no obvious defect throughout the layer. Note that in $Nd_{1-x}Sr_xNiO_2$, the pure perovskite and the resultant infinite-layer structure can only survive up to ~10 nm above which the Ruddlesden–Popper-type phase emerges (*11, 12, 34*). For the La-nickelate thin films, the pure infinite-layer phase is still observed for a thickness of 17 nm (Fig. 1B and Supplementary Data Fig. S1). We propose two possible explanations: (1) the lattice match between $La_{1-x}Ca_xNiO_3$ and $SrTiO_3$ substrate is much smaller than for the $Nd_{1-x}Sr_xNiO_3$, thus allowing better growth of the parent perovskite compound. (2) a more similar ionic size between $La^{3+}$ and the dopant ion $Ca^{2+}$, as compared to $Nd^{3+}$ vs $Sr^{2+}$. While the observation of a stable infinite-layer phase at a larger thickness in $La_{1-x}Ca_xNiO_2$ is still preliminary, it offers an alternative route for realizing bulk-like superconducting nickelates. In addition, while the $La_{1-x}Ca_xNiO_2$ was only recently studied, we notice high reproducibility in realizing superconductivity in the system as compared to $Nd_{1-x}Sr_xNiO_2$. Figure 1C shows the XRD $\theta$–$2\theta$ patterns of infinite-layer $La_{1-x}Ca_xNiO_2$ thin films with different Ca doping levels $x$ from 0 to 0.35. The (00$l$) peak positions slightly shift towards a higher angle as Ca dopant increases, indicating a shrinking of the $c$-axis lattice constants $d$ from ~3.405 Å at $x = 0$ to ~3.368 Å at $x = 0.35$ as plotted in Fig. 1D. The evolution of $d$ is in agreement with the empirical expectation as the cation is replaced with an atom having a smaller ionic radius. The Reciprocal space mappings around the (−103) diffraction peak indicate that the films are slightly relaxed, showing the larger in-plane lattice constants compared with that of $SrTiO_3$ substrate (Fig. 1D and Supplementary Data Fig. S2). The full width at half-maximum (FWHM) of the rocking curves for the (002) peaks of the infinite-layer films shows a value between 0.06° and 0.12°, supporting the high-quality thin-films after reduction (Supplementary Data Fig. S3).

**Electronic properties**

Figure 2A shows the logarithmic-scale resistivity versus temperature ($\rho$-$T$) curves of the La$_{1-x}$Ca$_x$NiO$_2$ thin films with Ca doping level $x$ from 0 to 0.35. For $x \leq 0.15$, the samples show insulating behavior all the way below 300 K. This is different from Nd$_{1-x}$Sr$_x$NiO$_2$ and Pr$_{1-x}$Sr$_x$NiO$_2$ thin films, in which the undoped and underdoped samples show a metallic behavior at high temperatures with a resistivity minimum at the intermediate temperature below which a weakly insulating behavior appears (*1, 11-13, 15*). The high-temperature metallic behavior in undoped Nd- and Pr-nickelates has been thought to be due to the self-doped nature of the parent compound (*2, 4, 35, 36*), however, theoretical simulations have not arrived at a dissimilar conclusion for La-nickelates (*4*). The zoomed-in and linear-scale $\rho$-$T$ curves of the superconducting La$_{1-x}$Ca$_x$NiO$_2$ thin films with $x$ from 0.15 to 0.27 are shown in Fig. 2B. For $x$ = 0.15 and 0.18, the samples show transitions from a slightly metallic behavior to an insulating behavior and then an onset of superconducting transition with decreasing temperature, while for $0.2 \leq x \leq 0.27$, the samples are superconducting with the $\rho$-$T$ curves showing metallic behavior at high temperatures. The suppression of superconductivity with increasing out-of-plane magnetic field for a representative sample with $x$ = 0.23 is shown in Supplementary Data Fig. S4. The fitting of the relationship between the upper critical field and mid-point critical temperature by the Ginzburg-Landau model gives the zero-temperature in-plane coherence length of 4.59 nm, comparable to that of $Nd_{0.8}Sr_{0.2}NiO_2$ film (*1*). At the overdoped regime with $x \geq 0.3$, the samples show an increase in normal state resistivity compared with that of the optimally doped sample. Moreover, similar to the Nd- and Pr-nickelates, the overdoped $La_{1-x}Ca_xNiO_2$ thin films exhibit weakly insulating behavior at low temperature, suggesting a universal transport property in over-doped infinite-layer nickelates (*11, 12, 15*).

Similar to Nd- and Pr-nickelates, the room-temperature $R_H$ is negative, and its magnitude decreases with increasing $x$ and then saturates at $x = 0.2$ (Fig. 2C and 2D), suggesting the multiband structure nature for $La_{1-x}Ca_xNiO_2$ films (*2, 4, 23, 24, 31, 35-37*). The $R_H$ remains negative below 300 K for samples with $x \leq 0.2$. For samples with $x \geq 0.23$ (except for $x = 0.25$), the $R_H$ undergoes a smooth transition from negative to positive sign as the temperature decreases. Figure 2D presents the $R_H$ at 20 K, clearly showing a sign change at $x = 0.2 - 0.23$ from negative to positive with increasing $x$. The doping level of the $R_H$-sign-change is higher than that ($x = 0.18 - 0.2$) of Nd- and Pr-nickelates (*11, 12, 15*). While we notice that the $R_H$-sign-change doping level for both La-nickelates and Nd-/Pr-nickelates are at around the optimally doped regime, in $Nd_{1-x}Sr_xNiO_2$ and $Pr_{1-x}Sr_xNiO_2$, the $R_H$-sign-change temperature increases with increasing doping level (*11, 12, 15*). However, for $La_{1-x}Ca_xNiO_2$ films, the $R_H$-sign-change temperature does not change with Ca doping level, and it is at around 35 K for $x \geq 0.23$. Such distinct saturation may suggest that hole doping in the superconducting lanthanide nickelates plays a more hidden role than merely charge carrier modulation.

**Phase diagram and superconducting dome**

Figure 3 depicts the phase diagrams of $La_{1-x}Ca_xNiO_2$ integrated with those of $Nd_{1-x}Sr_xNiO_2$ and $Pr_{1-x}Sr_xNiO_2$ (*11, 12, 15*). The onset of the critical temperature, $T_{c,90\%R}$, is defined as the temperature at which the resistivity drops to 90% of the normal-state value at the onset of the superconductivity. A superconducting dome between $0.15 < x < 0.3$ is seen for $La_{1-x}Ca_xNiO_2$. The doping range of the $La_{1-x}Ca_xNiO_2$ superconducting dome is comparable to that of $Pr_{1-x}Sr_xNiO_2$ and slightly wider than that of $Nd_{1-x}Sr_xNiO_2$ but is generally extended towards a higher doping level (Fig. 3B). At the two ends of the dome, the underdoped regime shows insulating behavior upon cooling, and the overdoped regime is weakly insulating at low temperature, different from the Nd- and Pr-nickelates in which both regimes show weakly insulating behavior (*11, 12, 15*). In general, the $T_c$ of $La_{1-x}Ca_xNiO_2$ is lower than Nd and Pr series, possibly due to the large ionic radius of La, consistent with the experimental observation in Cu- and Fe-based superconductors that the $T_c$ is enhanced as the ionic radius of the rare-earth element decreases (*29, 30*). This observation implies that the appearance of superconductivity in infinite layer nickelates is not reliant on the $4f$ electrons of the rare-earth element and suggests that higher $T_c$ in infinite-layer nickelates can be stabilized using smaller rare-earth ions.

**Discussion**

In addition to introducing charge carriers, the chemical doping in infinite-layer nickelate can cause the change of the multiband structures due to the modification of the lattice environment (*4, 31-33*). Therefore, the normal-state properties and the superconducting range will depend on the lattice constant upon doping, which can be seen from the doping dependent $R_H$ and superconducting dome (*11, 12, 15*). Depending on the difference of cation and dopant radius, the extent of the lattice-constant change upon doping is different. In $Nd_{1-x}Sr_xNiO_2$ (*11, 12*), the extent of the lattice-constant change with Sr doping is ~0.53 Å per one Sr atom, larger than that (~0.42 Å per one Sr atom) for $Pr_{1-x}Sr_xNiO_2$ (*15*) and that (~0.11 Å per one Ca atom) for $La_{1-x}Ca_xNiO_2$ that is observed in the current result. Interestingly, it is found that the superconducting dome of the $Nd_{1-x}Sr_xNiO_2$ is narrower than that of the $Pr_{1-x}Sr_xNiO_2$ and $La_{1-x}Ca_xNiO_2$, suggesting the intrinsic link between the superconducting range and the modulation of the lattice constant. Moreover, theoretical calculations suggested that Sr doping in $Nd_{1-x}Sr_xNiO_2$ reduces the self-doping effect, and the nickelate behaves more like a system with a pure single-band picture (*4, 33*). This is consistent with the $R_H$-sign-change upon doping, and the sign-change temperature monotonically increases with increasing Sr doping level in Nd- and Pr-nickelates (*11, 12, 15*). However, for $La_{1-x}Ca_xNiO_2$ films, the doping-dependent $R_H$-sign-change temperature is negligible for $x \geq 0.23$. Coupled with the small change in the lattice constant, this suggests the critical role of the dopant ionic nature on the electronic modulation in the infinite-layer nickelates. For the infinite-layer bulk nickelates $Nd_{1-x}Sr_xNiO_2$ and $Sm_{1-x}Sr_xNiO_2$ in which superconductivity is absent, the *c*-lattice constant is smaller than those of superconducting thin films (*38-40*). Moreover, we also found that the strain effect induced electronic bandstructure modulation and the resultant change of $R_H$ in $Nd_{0.8}Sr_{0.2}NiO_2$ (*41*). Recently, superconductivity was also reported in (La, Sr)$NiO_2$ (*42*). These suggest the importance of the lattice environment on the superconductivity and normal state properties in the infinite-layer nickelates. The present result also suggests a new route for new superconductors and a unique way for tailoring normal state properties through dopants in the nickelate superconductors.

**Materials and methods**

We prepared and characterized our Ca-doped LaNiO$_2$ samples following procedures previously described in [12].

**Sample preparation**

The ceramic targets with nominal composition La$_{1-x}$Ca$_x$NiO$_3$ were prepared by the conventional solid-state reaction using high-purity La$_2$O$_3$ (99.999%, Sigma-Aldrich), NiO$_2$ (99.99%, Sigma-Aldrich), CaCO$_3$ (99.995%, Sigma-Aldrich) powders as the starting materials. The mixed powders were sintered in air for 15 h at 1150, 1200 and 1250 °C, respectively, with thorough regrinding before each sintering. On the final sintering, the powder was pressed into a disk-shaped pellet. The perovskite thin films were grown on TiO$_2$-terminated (001) SrTiO$_3$ substrates using a pulsed laser deposition (PLD) technique with a 248-nm KrF excimer laser. The deposition temperature and oxygen partial pressure $P$O$_2$ for all samples were 600 °C and 150 mTorr, respectively. The laser energy density on the target surface was set to be 1.8 Jcm$^{-2}$. After deposition, the samples were annealed for 10 min at 600 °C and 150 mTorr and then cooled down to room temperature at a rate of 8 °C/min. In order to obtain the infinite-layer structures, the as-grown films were then embedded with about 0.15 g of CaH$_2$ powder and wrapped in aluminium foil, and then placed into the PLD chamber for the reduction process. The wrapped sample was heated to $340 - 360$°C at a rate of 25 °C/min and kept for 80 minutes, and then cooled down to room temperature at a rate of 25 °C/min.

**Electronic transport and structural characterization**

The transport measurements were performed using a Quantum Design Physical Property Measurement System. The wire connection for the electrical transport measurement was made by Al ultrasonic wire bonding. The X-ray diffraction (XRD) measurement was done in the X-ray Diffraction and Development (XDD) beamline at Singapore Synchrotron Light Source (SSLS) with an X-ray wavelength of $\lambda = 1.5404$ Å. The high-angle annular dark-field scanning transmission electron microscopy (HAADF-STEM) imaging was carried out at 200 kV using a JEOL ARM200F microscope, and the cross-sectional TEM specimens were prepared by a focused ion beam machine (FEI Versa 3D).

**Acknowledgments**

This research is supported by the Agency for Science, Technology, and Research (A*STAR) under its Advanced Manufacturing and Engineering (AME) Individual Research Grant (IRG) (A1983c0034). SWZ, CJL, YC and AA also acknowledge the partial support from the Singapore National Research Foundation (NRF) under the Competitive Research Programs (CRP Grant No. NRF-CRP15-2015-01). PY is supported by Singapore Synchrotron Light Source (SSLS) via NUS Core Support C-380-003-003-001. The authors would also like to acknowledge the SSLS for providing the facility necessary for conducting the research. The Laboratory is a National Research Infrastructure under the National Research Foundation (NRF) Singapore.

## Author contributions

SWZ and AA conceived the project. SWZ, LEC, YC, ZTZ, ZSL, JXH prepared the thin films and conducted the electrical measurements. SWZ, PY, CST, and XMY conducted the XRD measurements. CJL conducted the STEM measurements. SWZ and AA wrote the manuscript with contributions from all authors. All authors have discussed the results and the interpretations.

## Competing interests

All authors declare that they have no competing interests.

## Data and materials availability

All data needed to evaluate the conclusions in the paper are present in the paper and/or the Supplementary Materials.

## References

1. D. Li *et al.*, Superconductivity in an infinite-layer nickelate. *Nature* **572**, 624-627 (2019).
2. A. S. Botana, F. Bernardini, A. Cano, Nickelate superconductors: an ongoing dialog between theory and experiments. *Journal of Experimental and Theoretical Physics* ***132,*** *618–627* (2021).

3.  B. H. Goodge *et al.*, Doping evolution of the Mott–Hubbard landscape in infinite-layer nickelates. *Proceedings of the National Academy of Sciences* **118**, (2021).

4.  A. S. Botana, M. R. Norman, Similarities and Differences between LaNiO$_2$ and CaCuO$_2$ and Implications for Superconductivity. *Physical Review X* **10**, 011024 (2020).

5.  G. A. Sawatzky, Superconductivity seen in a non-magnetic nickel oxide. *Nature* **572**, (2019).

6.  M. Jiang, M. Berciu, G. A. Sawatzky, Critical Nature of the Ni Spin State in Doped NdNiO$_2$. *Physical Review Letters* **124**, 207004 (2020).

7.  S. Catalano *et al.*, Rare-earth nickelates RNiO$_3$: thin films and heterostructures. *Reports on Progress in Physics* **81**, 046501 (2018).

8.  M. A. G. M. A. Hayward, M. J. Rosseinsky, and J. Sloan, Sodium Hydride as a Powerful Reducing Agent for Topotactic Oxide Deintercalation-Synthesis and Characterization of the Nickel(I) Oxide LaNiO2. *J. Am. Chem. Soc.* **121**, 8843 (1999).

9.  J. G. Bednorz, K. A. Müller, Possible highT c superconductivity in the Ba– La– Cu– O system. *Zeitschrift für Physik B Condensed Matter* **64**, 189-193 (1986).

10. Y. Kamihara, T. Watanabe, M. Hirano, H. Hosono, Iron-based layered superconductor La[O$_{1-x}$F$_x$]FeAs (x = 0.05– 0.12) with T$_c$= 26 K. *Journal of the American Chemical Society* **130**, 3296-3297 (2008).

11. D. Li *et al.*, Superconducting dome in Nd$_{1-x}$Sr$_x$NiO$_2$ infinite layer films. *Physical Review Letters* **125**, 027001 (2020).

12. S. Zeng *et al.*, Phase diagram and superconducting dome of infinite-layer Nd$_{1-x}$Sr$_x$NiO$_2$ thin films. *Physical Review Letters* **125**, 147003 (2020).

13. M. Osada *et al.*, A Superconducting Praseodymium Nickelate with Infinite Layer Structure. *Nano Lett* **20**, 5735-5740 (2020).

14. Q. Gu *et al.*, Single particle tunneling spectrum of superconducting Nd$_{1-x}$Sr$_x$NiO$_2$ thin films. *Nature communications* **11**, 1-7 (2020).

15. M. Osada, B. Y. Wang, K. Lee, D. Li, H. Y. Hwang, Phase diagram of infinite layer praseodymium nickelate $Pr_{1-x}Sr_xNiO_2$ thin films. *Physical Review Materials* **4**, 121801 (2020).

16. Q. Gao, Y. Zhao, X. Zhou, Z. Zhu, Preparation of superconducting thin film of infinite-layer nickelate $Nd_{0.8}Sr_{0.2}NiO_2$. *arXiv preprint arXiv:2102.10292*, (2021).

17. X. Zhou *et al.*, Negligible oxygen vacancies, low critical current density, electric-field modulation, in-plane anisotropic and high-field transport of a superconducting $Nd_{0.8}Sr_{0.2}NiO_2/SrTiO_3$ heterostructure. *arXiv preprint arXiv:2104.07316*, (2021).

18. B. Y. Wang *et al.*, Isotropic Pauli-limited superconductivity in the infinite-layer nickelate $Nd_{0.775}Sr_{0.225}NiO_2$. *Nature Physics*, 1-5 (2021).

19. H. Lu *et al.*, Magnetic excitations in infinite-layer nickelates. *arXiv preprint arXiv:2105.11300*, (2021).

20. P. Puphal *et al.*, Synthesis and Characterization of Ca-Substituted Infinite-Layer Nickelate Crystals. *arXiv preprint arXiv:2106.13171*, (2021).

21. J. Zhang *et al.*, Large orbital polarization in a metallic square-planar nickelate. *Nature Physics* **13**, 864-869 (2017).

22. L. Si *et al.*, Topotactic Hydrogen in Nickelate Superconductors and Akin Infinite-Layer Oxides $ABO_3$. *Physical Review Letters* **124**, 166402 (2020).

23. P. Jiang, L. Si, Z. Liao, Z. Zhong, Electronic structure of rare-earth infinite-layer $RNiO_2$ (R= La, Nd). *Physical Review B* **100**, 201106(R) (2019).

24. M.-Y. Choi, K.-W. Lee, W. E. Pickett, Role of 4f states in infinite-layer $NdNiO_2$. *Physical Review B* **101**, 020503(R) (2020).

25. S. Bandyopadhyay, P. Adhikary, T. Das, I. Dasgupta, T. Saha-Dasgupta, Superconductivity in infinite-layer nickelates: Role of f orbitals. *Physical Review B* **102**, 220502 (2020).

26. Y. Nomura, M. Hirayama, T. Tadano, Y. Yoshimoto, K. Nakamura, R. Arita, Formation of a two-dimensional single-component correlated electron system and band engineering in the nickelate superconductor NdNiO$_2$. *Physical Review B* 100, 205138 (2019).

27. X. Wu, D. Di Sante, T. Schwemmer, W. Hanke, H. Y. Hwang, S. Raghu, R. Thomale, Robust $d_{x^2-y^2}$-wave superconductivity of infinite-layer nickelates. *Physical Review B* **101**, 060504(R) (2020).

28. Z. Liu *et al.*, Doping dependence of electronic structure of infinite-layer NdNiO$_2$. *Physical Review B* **103**, 045103 (2021).

29. H. Hosono, K. Kuroki, Iron-based superconductors: Current status of materials and pairing mechanism. *Physica C: Superconductivity and its Applications* **514**, 399-422 (2015).

30. N. P. Armitage, P. Fournier, R. L. Greene, Progress and perspectives on electron-doped cuprates. *Reviews of Modern Physics* **82**, 2421-2487 (2010).

31. J. Gao, Z. Wang, C. Fang, H. Weng, Electronic structures and topological properties in nickelates $Ln_{n+1}Ni_nO_{2n+2}$. *National Science Review* **8** (2021).

32. F. Lechermann, Doping-dependent character and possible magnetic ordering of NdNiO$_2$. *Physical Review Materials* **5**, 044803 (2021).

33. M. Kitatani *et al.*, Nickelate superconductors—a renaissance of the one-band Hubbard model. *npj Quantum Materials* **5**, 1-6 (2020).

34. K. Lee *et al.*, Aspects of the synthesis of thin film superconducting infinite-layer nickelates. *APL Materials* **8**, 041107 (2020).

35. G.-M. Zhang, Y.-f. Yang, F.-C. Zhang, Self-doped Mott insulator for parent compounds of nickelate superconductors. *Physical Review B* **101**, 020501(R) (2020).

36. M. Hepting *et al.*, Electronic structure of the parent compound of superconducting infinite-layer nickelates. *Nat Mater* **19**, 381-385 (2020).

37. K.-W. Lee, W. Pickett, Infinite-layer LaNiO$_2$: Ni$^{1+}$ is not Cu$^{2+}$. *Physical Review B* **70**, 165109 (2004).

38. Q. Li *et al.*, Absence of superconductivity in bulk Nd$_{1-x}$Sr$_x$NiO$_2$. *Communications Materials* **1**, 16 (2020).

39. B.-X. Wang *et al.*, Synthesis and characterization of bulk $Nd_{1-x}Sr_xNiO_2$ and $Nd_{1-x}Sr_xNiO_3$. Physical Review Materials **4**, 084409 (2020).

40. C. He *et al.*, Synthesis and physical properties of perovskite $Sm_{1-x}Sr_xNiO_3$ (x = 0, 0.2) and infinite-layer $Sm_{1-x}Sr_xNiO_2$ nickelates. *J Phys Condens Matter*, (2021).

41. S. Zeng *et al.*, Observation of perfect diamagnetism and interfacial effect on the electronic structures in $Nd_{0.8}Sr_{0.2}NiO_2$ superconducting infinite layers. *arXiv preprint arXiv:2104.14195*, (2021).

42. M. Osada *et al.*, Nickelate superconductivity without rare-earth magnetism:(La, Sr)$NiO_2$. *arXiv preprint arXiv:2105.13494*, (2021).

**Figure Captions**

**Figure 1| Structural characterization of infinite-layer $La_{1-x}Ca_xNiO_2$ thin films.** (**A**) The XRD $\theta$–$2\theta$ scan patterns of the perovskite $La_{0.77}Ca_{0.23}NiO_3$ and infinite-layer $La_{0.77}Ca_{0.23}NiO_2$ thin films. The arrow denotes the transition of diffraction peaks related to the transformation from a perovskite to an infinite-layer structure. (**B**) The HAADF-STEM image of the 17-nm $La_{0.77}Ca_{0.23}NiO_2$ on $SrTiO_3$ substrate. (**C**) The XRD $\theta$–$2\theta$ scan patterns of the $La_{1-x}Ca_xNiO_2$ thin films with different Ca doping levels $x$. The intensity is vertically displaced for clarity. (**D**) The room-temperature $c$-axis lattice constants, $d$, and in-plane lattice constant, $a$, as a function of $x$.

**Figure 2| Electrical transport properties of infinite-layer $La_{1-x}Ca_xNiO_2$ thin films.** (**A**) The logarithmic-scale resistivity versus temperature ($\rho$-$T$) curves of the $La_{1-x}Ca_xNiO_2$ thin films with Ca doping level $x$ from 0 to 0.35, measured from 300 to 2 K. (**B**) The zoomed-in and linear-scale $\rho$-$T$ curves of the $La_{1-x}Ca_xNiO_2$ thin films with $x$

from 0.15 to 0.27. For clarity, the resistivity of samples with $x$ = 0.15, 0.18 and 0.27 is diminished to 0.3 of the initial value. (**C**) The temperature dependence of the normal-state Hall coefficients $R_H$. (**D**) The $R_H$ at $T$ = 300 and 20 K as a function of $x$. The dash lines are guides to the eye.

**Figure 3| Phase diagram of infinite-layer La$_{1-x}$Ca$_x$NiO$_2$ thin films and the comparison to Nd$_{1-x}$Sr$_x$NiO$_2$ and Pr$_{1-x}$Sr$_x$NiO$_2$ thin films.** (**A**) The critical temperature as a function of doping level $x$ for La$_{1-x}$Ca$_x$NiO$_2$ in the present results. The $T_{c,90\%R}$ is defined as the temperature at which the resistivity drops to 90% of the value at 10 K (the onset of the superconductivity). (**B**) The combined phase diagram of La$_{1-x}$Ca$_x$NiO$_2$, Nd$_{1-x}$Sr$_x$NiO$_2$ and Pr$_{1-x}$Sr$_x$NiO$_2$. The data of Nd$_{1-x}$Sr$_x$NiO$_2$ is adapted from references [11, 12], and the data of Pr$_{1-x}$Sr$_x$NiO$_2$ is adapted from reference [15].

**Figures and captions**

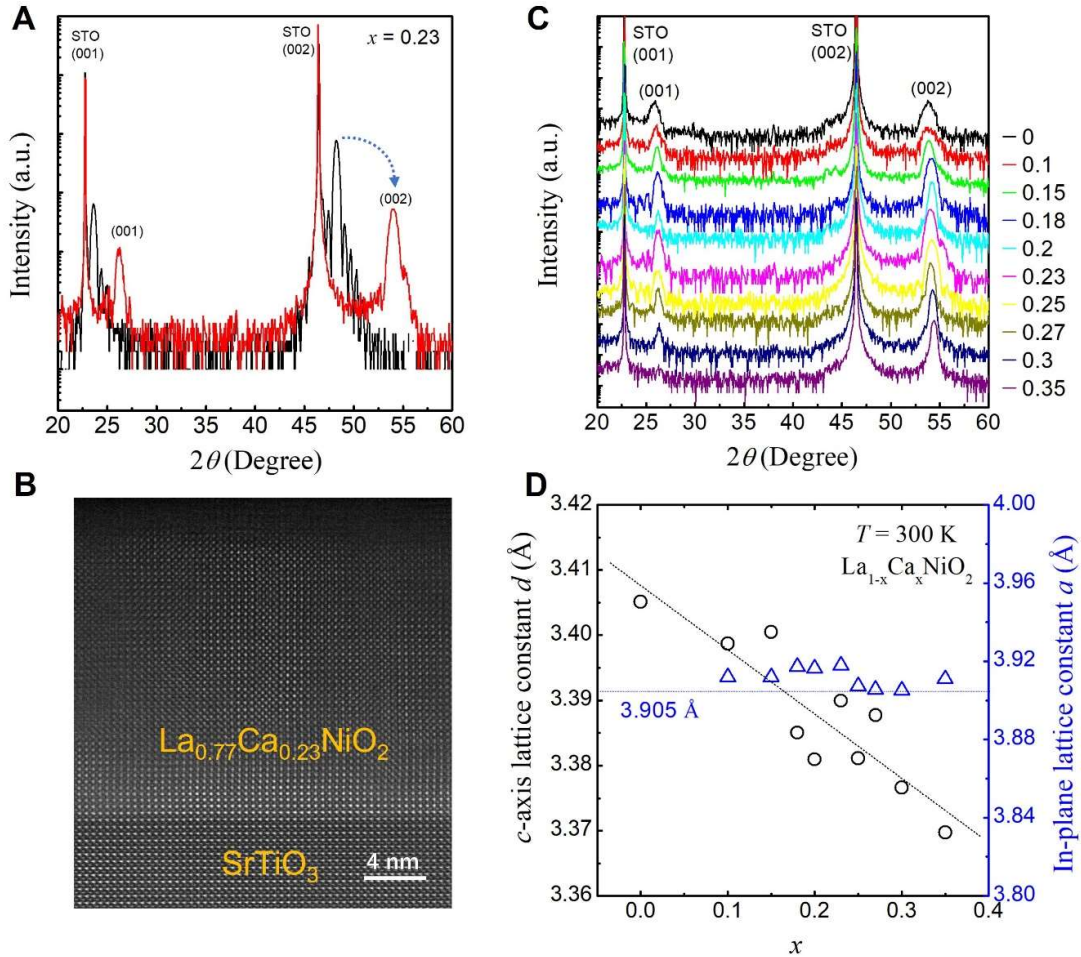

**Figure 1| Structural characterization of infinite-layer La$_{1-x}$Ca$_x$NiO$_2$ thin films.** (**A**) The XRD $\theta-2\theta$ scan patterns of the perovskite La$_{0.77}$Ca$_{0.23}$NiO$_3$ and infinite-layer La$_{0.77}$Ca$_{0.23}$NiO$_2$ thin films. The arrow denotes the transition of diffraction peaks related to the transformation from a perovskite to an infinite-layer structure. (**B**) The HAADF-STEM image of the 17-nm La$_{0.77}$Ca$_{0.23}$NiO$_2$ on SrTiO$_3$ substrate. (**C**) The XRD $\theta-2\theta$ scan patterns of the La$_{1-x}$Ca$_x$NiO$_2$ thin films with different Ca doping levels $x$. The intensity is vertically displaced for clarity. (**D**) The room-temperature $c$-axis lattice constants, $d$, and in-plane lattice constant, $a$, as a function of $x$.

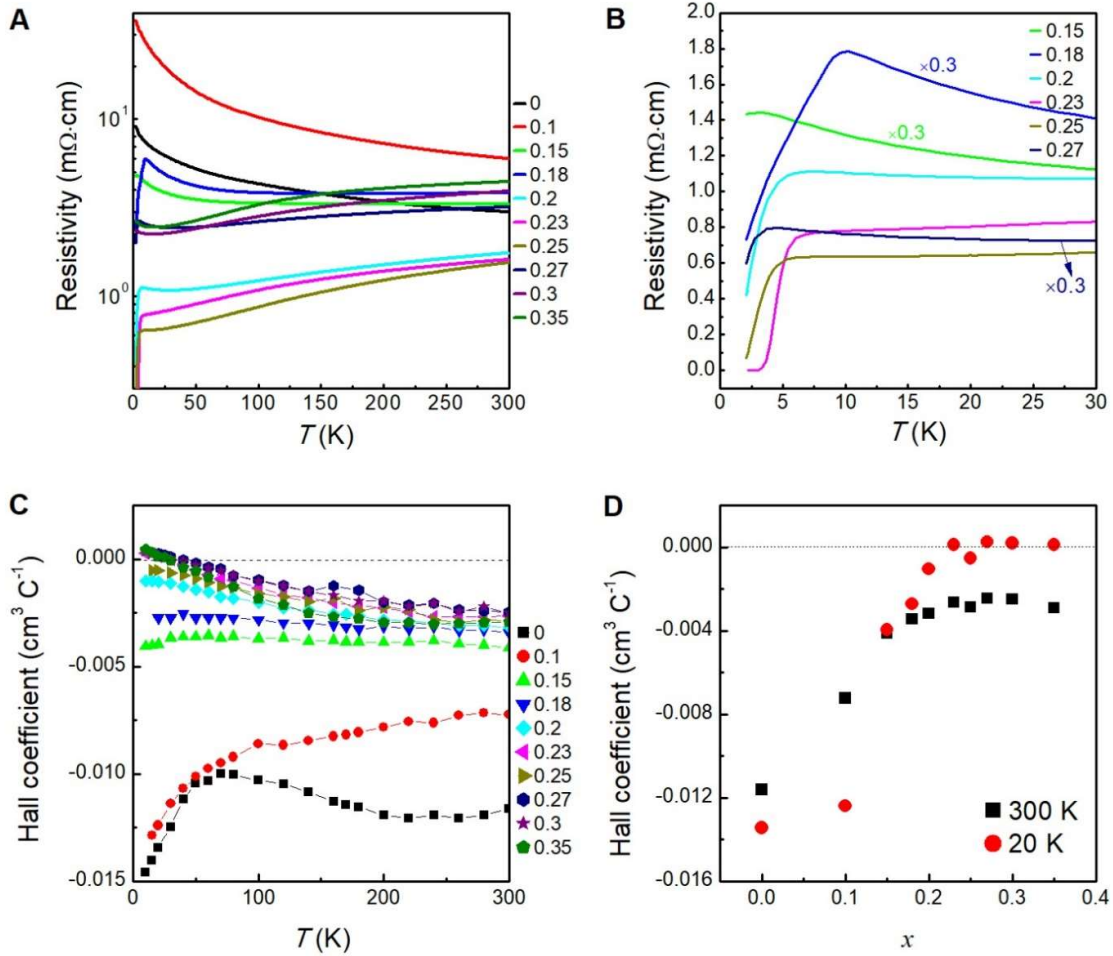

**Figure 2| Electrical transport properties of infinite-layer La$_{1-x}$Ca$_x$NiO$_2$ thin films.** (**A**) The logarithmic-scale resistivity versus temperature ($\rho$-$T$) curves of the La$_{1-x}$Ca$_x$NiO$_2$ thin films with Ca doping level $x$ from 0 to 0.35, measured from 300 to 2 K. (**B**) The zoomed-in and linear-scale $\rho$-$T$ curves of the La$_{1-x}$Ca$_x$NiO$_2$ thin films with $x$ from 0.15 to 0.27. For clarity, the resistivity of samples with $x = 0.15$, 0.18 and 0.27 is diminished to 0.3 of the initial value. (**C**) The temperature dependence of the normal-state Hall coefficients $R_H$. (**D**) The $R_H$ at $T = 300$ and 20 K as a function of $x$. The dash lines are guides to the eye.

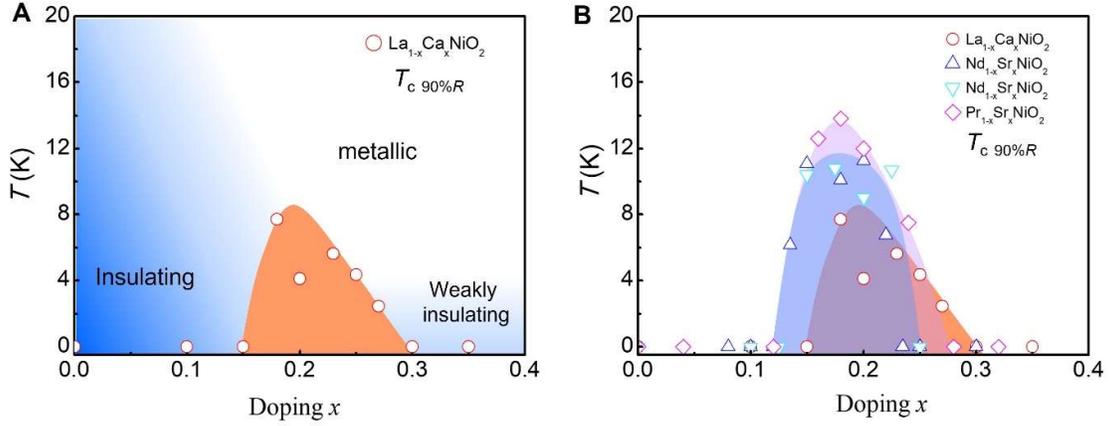

**Figure 3| Phase diagram of infinite-layer La$_{1-x}$Ca$_x$NiO$_2$ thin films and the comparison to Nd$_{1-x}$Sr$_x$NiO$_2$ and Pr$_{1-x}$Sr$_x$NiO$_2$ thin films.** (**A**) The critical temperature as a function of doping level $x$ for La$_{1-x}$Ca$_x$NiO$_2$ in the present results. The $T_{c,90\%R}$ is defined as the temperature at which the resistivity drops to 90% of the value at 10 K (the onset of the superconductivity). (**B**) The combined phase diagram of La$_{1-x}$Ca$_x$NiO$_2$, Nd$_{1-x}$Sr$_x$NiO$_2$ and Pr$_{1-x}$Sr$_x$NiO$_2$. The data of Nd$_{1-x}$Sr$_x$NiO$_2$ is adapted from references [11, 12], and the data of Pr$_{1-x}$Sr$_x$NiO$_2$ is adapted from reference [15].

# Supplementary Materials for

**Superconductivity in infinite-layer nickelate La$_{1-x}$Ca$_x$NiO$_2$ thin films**

S. W. Zeng[1,#,*], C. J. Li[2,3,#], L. E. Chow[1,#], Y. Cao[4], Z. T. Zhang[1], C. S. Tang[5,6], X. M. Yin[7], Z. S. Lim[1], J. X. Hu[1], P. Yang[2,5], A. Ariando[1,*]

*Corresponding author. Email: ariando@nus.edu.sg, shengwei_zeng@u.nus.edu

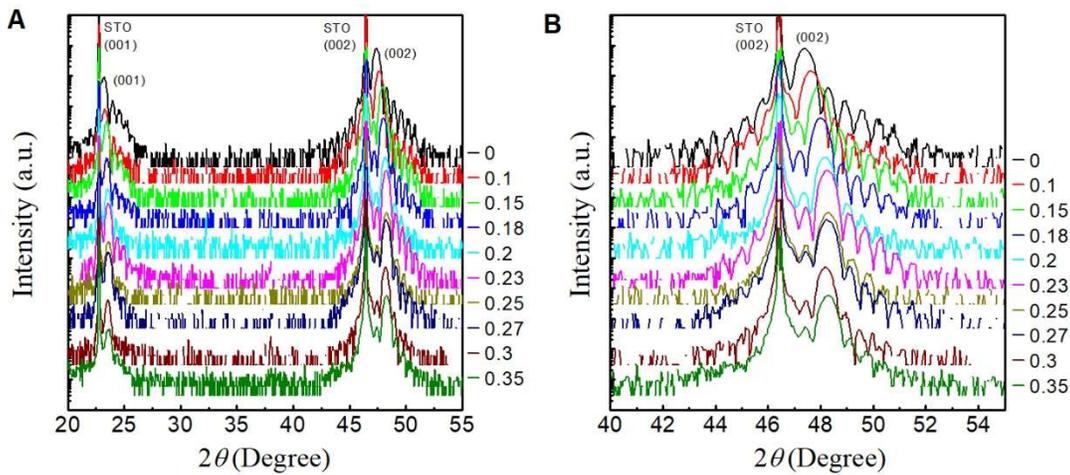

**Supplementary Data - Figure S1| The XRD $\theta - 2\theta$ scan of infinite-layer La$_{1-x}$Ca$_x$NiO$_2$ thin films.** (**A**) The XRD $\theta - 2\theta$ scan patterns of the as-grown perovskite La$_{1-x}$Ca$_x$NiO$_3$ thin films with different Ca doping levels $x$ from 0 to 0.35. The intensity is vertically displaced for clarity. Only the (00$l$) perovskite peaks are observed, where $l$ is an integer, confirming the $c$-axis oriented epitaxial growth. (**B**) The zoomed-in XRD $\theta$– $2\theta$ scan patterns at angles from 40 to 55 degrees. The clear thickness oscillations in the vicinity of the (002) peak indicate the single-phase and high quality of the perovskite

films. The thickness of around 17 nm is obtained by calculating the period of thickness oscillations in the vicinity of the (002) peak. The (002) peak positions shift towards a higher angle as the Ca doping level increase, indicating a shrinking of the $c$-axis lattice, in agreement with the empirical expectation when replacing the cation with an atom having a smaller ionic radius.

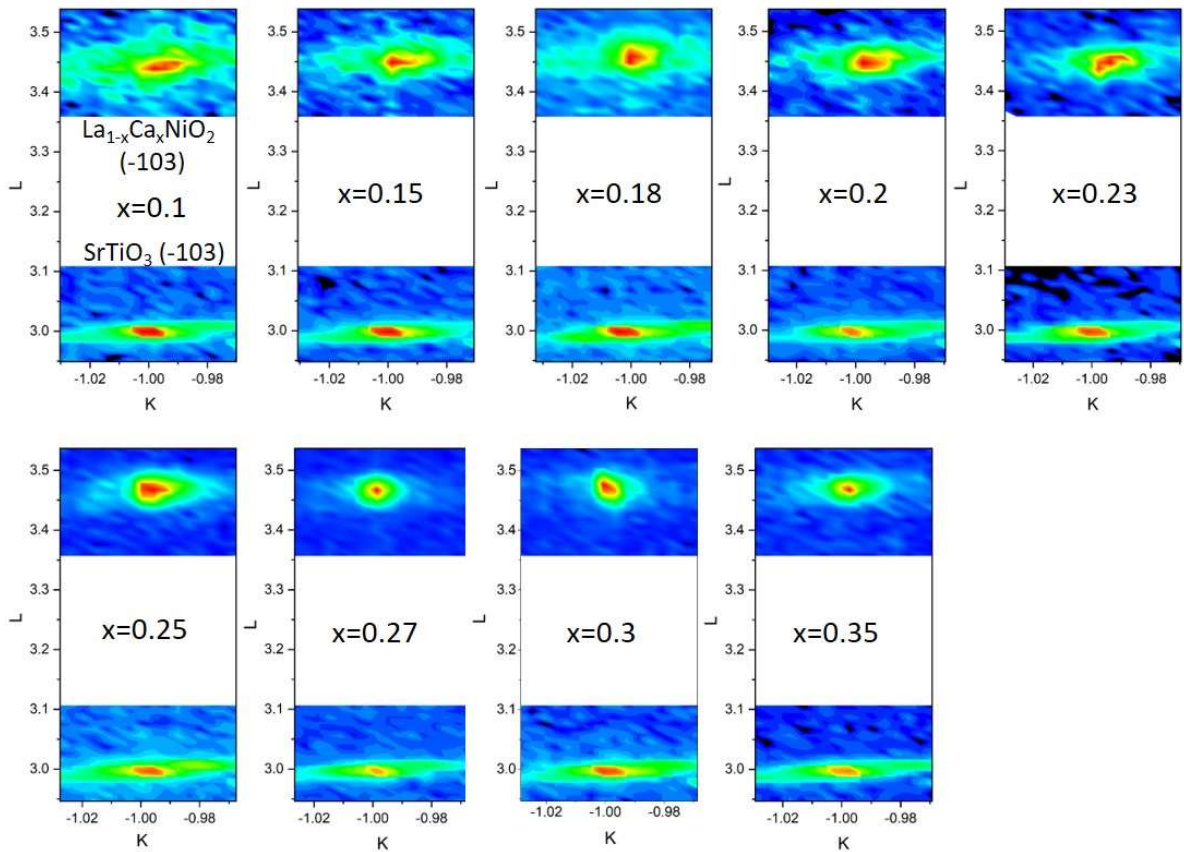

**Supplementary Data - Figure S2| The reciprocal space mapping scan of infinite-layer $La_{1-x}Ca_xNiO_2$ thin films.** The reciprocal space mapping scan is taken around ($\bar{1}03$) diffraction peak for $La_{1-x}Ca_xNiO_2$ thin films with Ca doping level $x$ from 0.1 to 0.35.

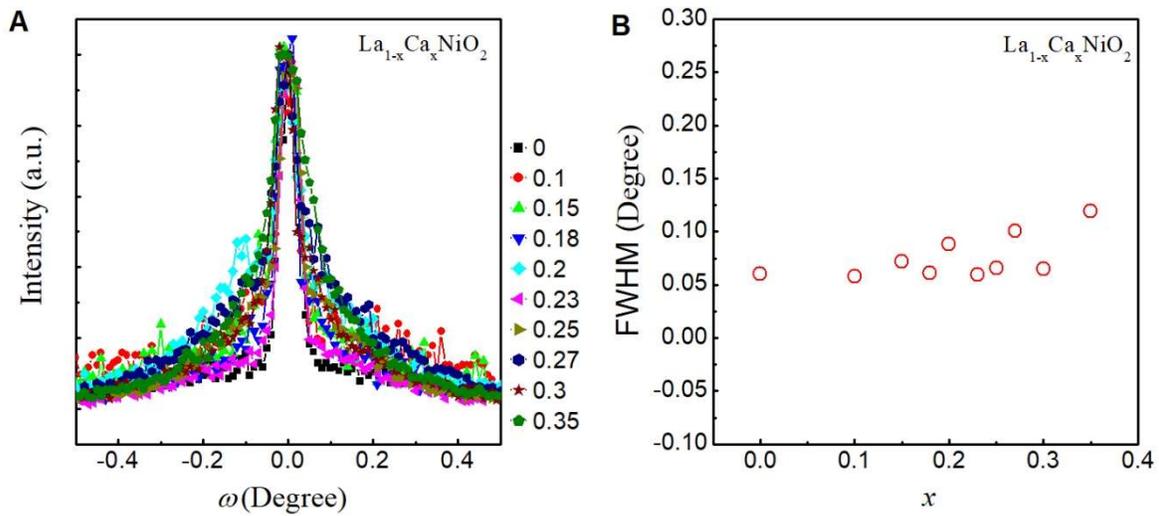

**Supplementary Data – Figure S3| The rocking curves of infinite-layer $La_{1-x}Ca_xNiO_2$ thin films.** (**A**) The rocking curves for the (002) peaks of the infinite-layer $La_{1-x}Ca_xNiO_2$ thin films with different Ca doping level $x$. (**B**) The full width at half-maximum (FWHM) of the (002) rocking curves as a function of $x$. The value of FWHM is between 0.06º and 0.12º.

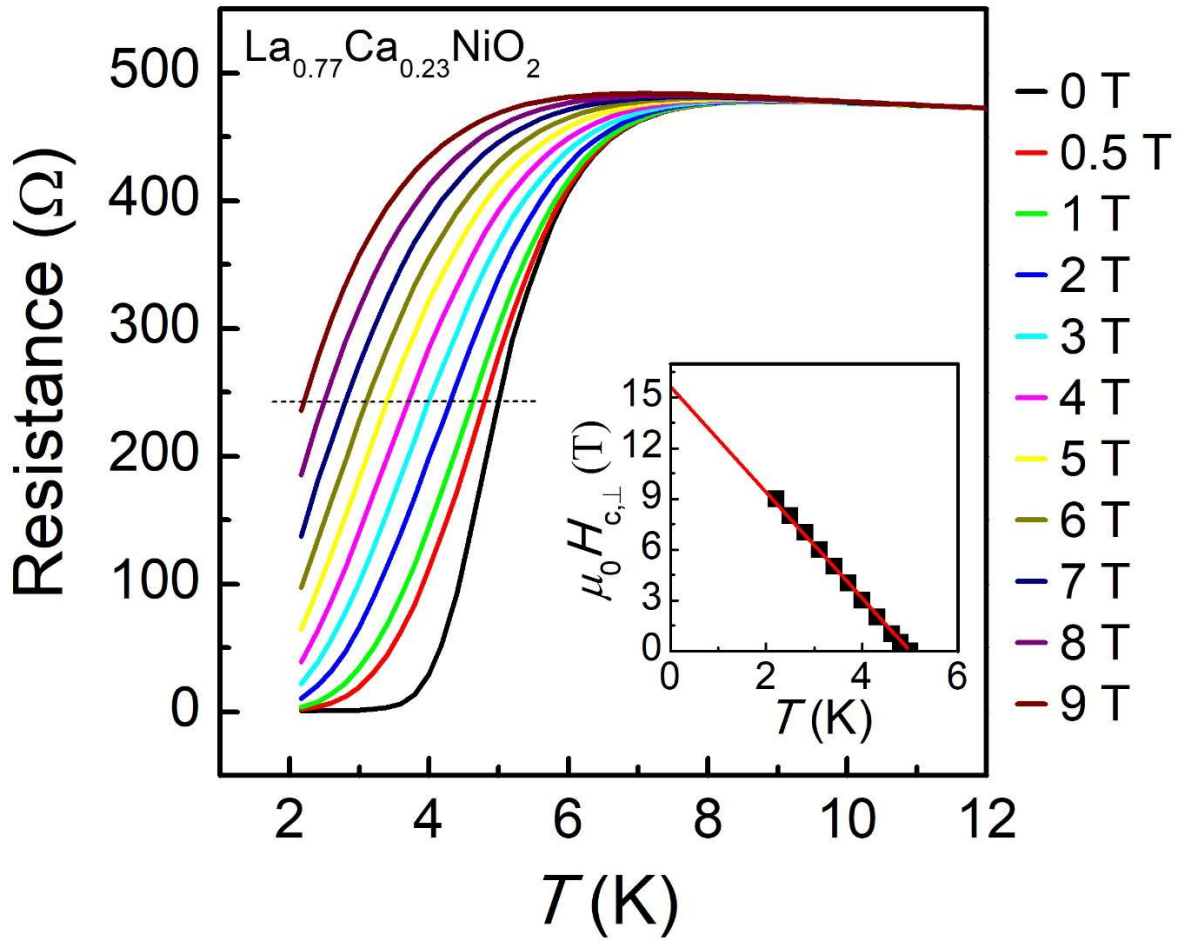

**Supplementary Data – Figure S4| Magnetic field dependence of the transition temperature.** The resistance-temperature (*R-T*) curves for a sample with Ca doping level of 0.23 under various magnetic fields applied perpendicularly to the *a-b* plane. The inset shows the relationship between the upper critical field $\mu_0 H_{c,\perp}$ and $T_C$ (extracted by the midpoint of the resistive transition) with a linear Ginzburg-Landau fitting. The red line is the fitting curve. The fitting gives a zero-temperature in-plane Ginzburg-Landau coherence length of 4.59 nm.

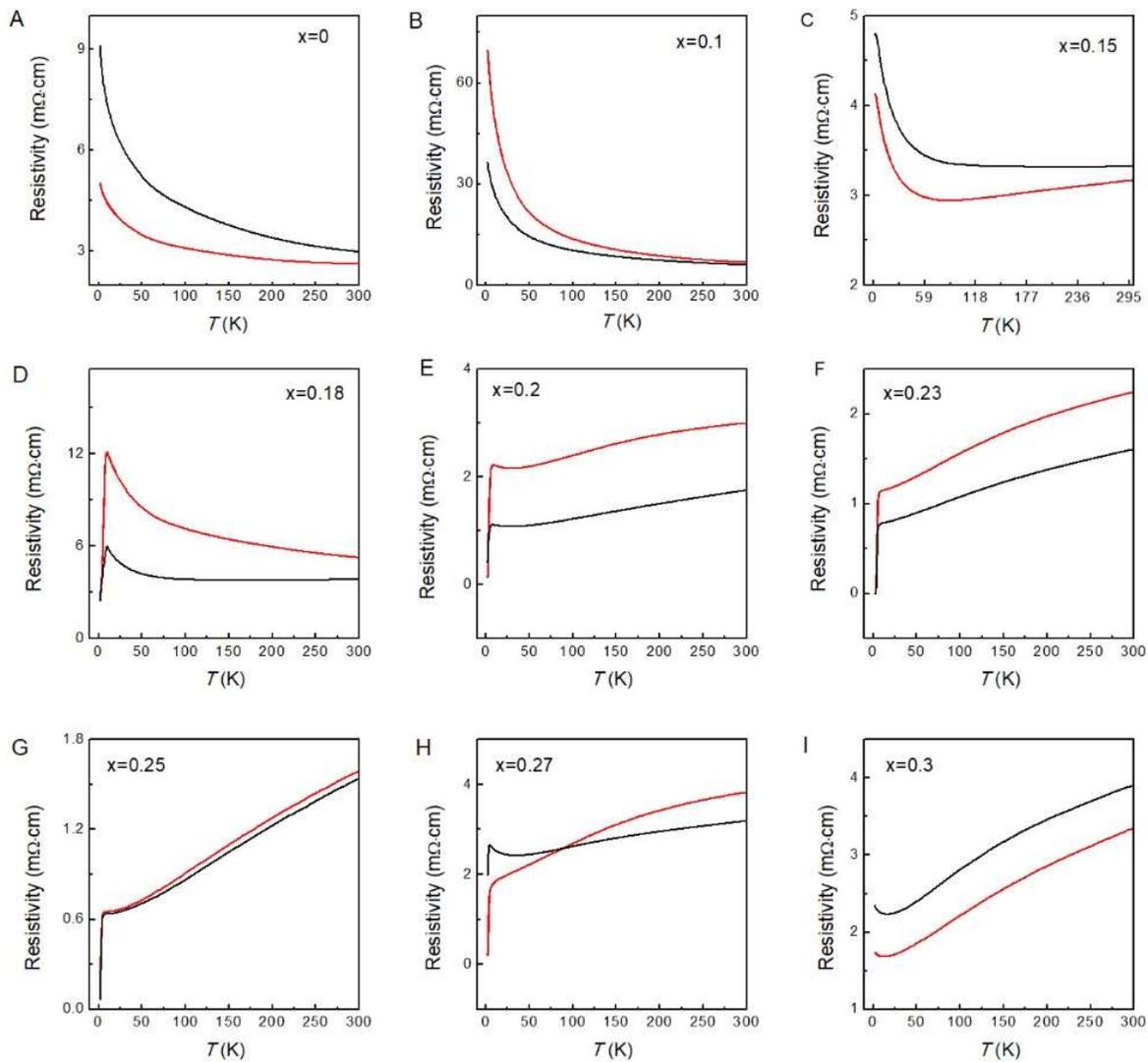

**Supplementary Data – Figure S5| Reproducibility of the transition temperature.** The resistivity-temperature (ρ-*T*) curves for two set of samples with Ca doping level varies from x=0 to x=0.3.

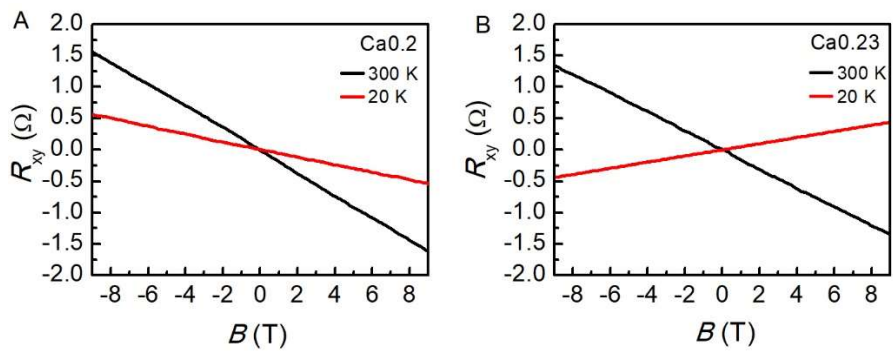

**Supplementary Data – Figure S6| Linear Hall resistance.** Hall resistance measured as a function of magnetic field at 300K vs 20K for (A) 0.2 Ca doping, and (B) 0.23 Ca doping.